\documentclass[%
superscriptaddress,
 amsmath,amssymb,
 aps,prfluids,
]{revtex4-2}
\usepackage{color, float}
\usepackage{graphicx}
\usepackage{dcolumn}
\usepackage{bm}
\usepackage{appendix}

\begin{document}

\preprint{APS/123-QED}

\title{Extended Spreading of Saline Droplets upon Impact on a Frosty Surface}


\author{Hao Zeng}
\affiliation{New Cornerstone Science Laboratory, Center for Combustion Energy, Key Laboratory for Thermal Science and Power Engineering of Ministry of Education,
Department of Energy and Power Engineering, Tsinghua University, 100084 Beijing, China}

\author{Feng Wang}
\thanks{fengwang2023@tsinghua.edu.cn}
\affiliation{New Cornerstone Science Laboratory, Center for Combustion Energy, Key Laboratory for Thermal Science and Power Engineering of Ministry of Education,
Department of Energy and Power Engineering, Tsinghua University, 100084 Beijing, China}

\author{Chao Sun}
 \thanks{chaosun@tsinghua.edu.cn}
\affiliation{New Cornerstone Science Laboratory, Center for Combustion Energy, Key Laboratory for Thermal Science and Power Engineering of Ministry of Education,
Department of Energy and Power Engineering, Tsinghua University, 100084 Beijing, China}
\affiliation{Department of Engineering Mechanics, School of Aerospace Engineering, Tsinghua University, Beijing 100084, China
}
\affiliation{Physics of Fluids Group, Max Planck Center for Complex Fluid Dynamics, and J.M.Burgers Center for Fluid Dynamics, University of Twente, 7500 AE Enschede, The Netherlands
}

\date{\today}

\begin{abstract}
Understanding the solidification dynamics of impacted water droplets is fundamental and crucial for applications, especially with the presence of frost and salt. Here, we experimentally investigate the spreading and freezing dynamics of saline droplets upon impact on a cold, frosty surface.
Our findings demonstrate that the frost and salt can lead to an extended spreading of impacted droplets under specific conditions.
In addition to the well-known 1/2 inertial spreading scaling law for droplets impact on a cold substrate, we observe a distinct transition to a 1/10 scaling law dominated by the capillary-viscous relation at low impacting velocities. 
We formulate the onset criterion for this extended spreading regime and derive a scaling dependence that captures the droplet's arrested diameter over various supercooling temperatures, by incorporating the effect of impact inertia, partial-wetting behavior and salinity.
 Based on the analysis, a unified model is proposed for predicting the droplet arrested diameter over a wide range of impact velocity and salinity.
\end{abstract}

\maketitle

\section{Introduction}

Ice accretion on natural and artificial surfaces can pose serious challenges to maritime operations in cold region. Extensive efforts have been made by researchers to comprehend the icing mechanisms and to devise anti-icing techniques to prevent icing and frosting \cite{stairs1971changes,thievenaz2020retraction,ghabache2016frozen,cao2009anti,ruan2013preparation,wan2023freezing,lyu2023liquid}. However, there are numerous inquiries regarding the mechanism of surface icing that need to be further investigated \cite{wildeman2017fast,meijer2023thin,wang2024self,zeng2023evaporation}, and some of these questions are still under debate \cite{kalita2023microstructure,grivet2022contact,kreder2016design,roisman2021wetting}. Most of the current studies focus on the pure water cases \cite{zeng2022influence,marin2014universality,fang2022self}, while the dissolved salt in the droplet has shown a growing importance when considering icing problems. From a geophysical standpoint, the freezing process of saline water is crucial for understanding the sea ice formation \cite{du2023sea}, ocean circulation \cite{shcherbina2003direct}, and the consequent climate problems \cite{clark1999northern,demott2016sea}. From an industrial applications perspective, sea spray icing can bring severe consequences to marine vessels and offshore structures for their stability, maneuverability, and overall safety \cite{dehghani2017sea,ryerson2011ice,lam2020effect}, and the rejection of dissolved salt even increases the risks on structural integrity and damage.


Although the anti-icing ability of salt has been demonstrated regarding to its freezing point depression property, the solidification process of saltwater droplets is intricate and encompasses different physical processes such as fluid flow, heat transfer, salt transport, and phase transition across a wide range of spatial and temporal scales. The influence of salinity on the solidification process of saline droplets has been investigated in numerous studies \cite{vrbka2005brine,bauerecker2008monitoring,singha2018influence,boinovich2016anti}: a considerable time delay for heterogeneous ice nucleation and brine rejection near the water-ice interface have been identified during the solidification process by experiments and molecular dynamic simulations. Though it is important in many branches of science and technology, the influence of kinetic effect is rarely considered in these studies. Carpenter \textit{et al.} investigated the impact of saline droplets on a cold superhydrophobic surface \cite{carpenter2015saltwater}. They observed a good repellency of the superhydrophobic surface to impact icing with saltwater for the surface temperature down to $-40$ $^\circ$C due to the absence of ice nucleus and the slowed ice nucleation propagation kinetics. However, when the solid substrate is the same material as the droplet, the nucleation energy barrier can be entirely diminished. As a result, in many application scenarios, the presence of frost on top of anti-icing surfaces can lead to significant icephobicity failure \cite{kreder2016design,lambley2023freezing,lolla2022arrested}. 
Previous studies have focused on how salt concentration affects ice nucleation and subsequent ice front propagation within droplets. Nevertheless, the presence of frost and dynamic impact process in many natural scenarios adds complexity to the process of saline droplet freezing. The question remaining is: What happens when a saline droplet encounters a cold, frosty surface? In this study, we attempt to answer it.

In this paper, we experimentally investigate the impact of saline droplets on a cold, frosty substrate.
Our results suggest that, as compared to pure water cases, the dissolved salt can lead to an extended spreading stage under certain conditions. Furthermore, a distinct transition from the $1/2$ inertial spreading power law to a $1/10$ capillary-viscous spreading is observed at low impact inertia ($We < 100$). We quantify the onset criterion and the arrested condition of the extended spreading regime, by considering the freezing point $T_f$ depression effect induced by droplet salinity $S_i$. Lastly, we propose a  model for the final arrested diameter across a broad range of impacting inertia and salinity.



\section{Experiment methods}
\subsection{Experiment setup}
\begin{figure*}[htbp]
\includegraphics[width=0.6\textwidth]{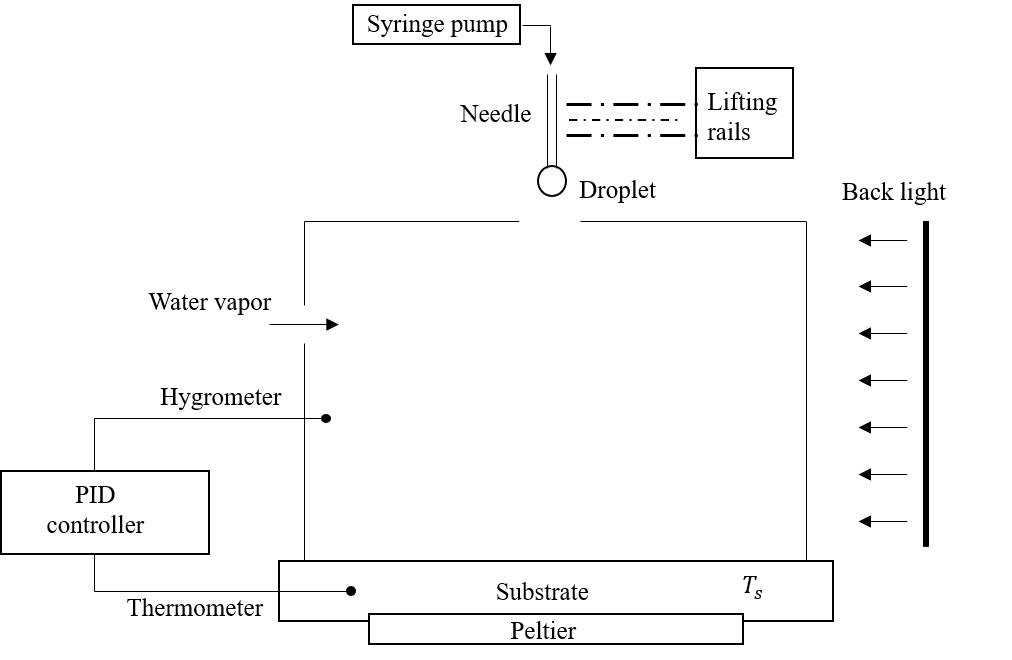} 
\caption{\label{fig:setup} Experimental setup.}
\end{figure*}

The experiments are conducted in a self-built transparent environmental chamber with an inner size of 8 cm $\times$ 8 cm $\times$ 8 cm, where the air humidity can be controlled and the environment temperature are monitored, as shown in Fig.~\ref{fig:setup}. In experiments, the liquid droplets are either pure water or salt solutions prepared by using sodium chloride supplied by Sigma Aldrich (ACS reagent, $\geq$ 99$\%$) and distilled water. These micro-liter droplets generated with a syringe pump and stainless steel needle with an initial diameter $D_0=3.4 \pm 0.25$ mm are released from different heights at room temperature to impact on a cooled nickel-plated copper substrate.

The surface of the substrate is kept at a constant temperature ($T_s=-15 ^{\circ}$C and $-20 ^{\circ}$C) by an embedded peltier cooler. A heat exchanger connecting to a cold reservoir is attached to the opposite side of the peltier cooler to maintain energy balance. The surface temperature is measured with a thermocouple embedded 0.5 mm beneath the top surface of the substrate. The thermocouple is connected with a PID controller to adjust the power supply to the cooler so that we can obtain a constant surface temperature with a variation of less than 0.2 K. 

The freezing process is recorded by a high-speed imaging system from the side view with back-light illumination. The imaging system utilizes a high-speed camera (Photron Nova S12), combined with a long-distance microscope (Zoom 6000 by Navitar), resulting in a maximum resolution of 5.7 $\mu$m/pixel. With this imaging system, we can precisely acquire high-resolution images for analyzing the droplet freezing features versus time. The evolution of droplet contact diameter after impact is measured using a self-developed Matlab code.

\subsection{Frost growth and characterization}
In the experiment, the ambient temperature in the environmental chamber is controlled by the PID system to a constant value $T_{\infty}=23~^{\circ}$C. At the beginning of each experiment, the chamber is flushed with pure nitrogen gas. To grow frost, the relative humidity is set to $H_{rel}=60\%$ by injecting water vapor. The frost growth is monitored from the top via microscopy and the frost growth time is set to be the same. Once the frost sheet was grown to the desired thickness, the water vapor injection was lowered to the minimum and then the droplet is released from the needle. This ensures a reproducible frost layer between each independent impact experiment.

\begin{figure*}[htbp]
\includegraphics[width=0.6\textwidth]{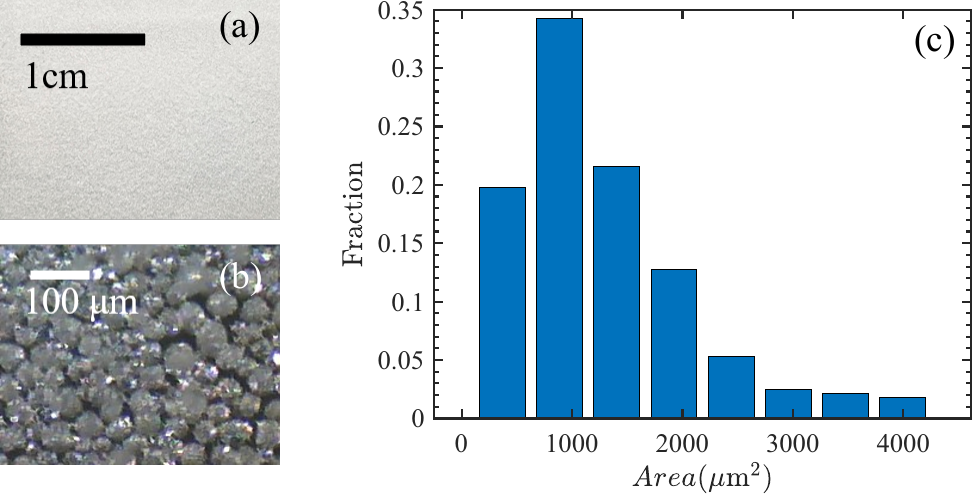} 
\caption{\label{fig:frost} Characterization of the frosty substrate.}
\end{figure*}

Fig.~\ref{fig:frost} shows the characterization of the frost growth on top of the chilled substrate. As shown in Fig.~\ref{fig:frost}(a), the frost is homogeneous in terms of appearance, since the supercooled filmwise condensation promoted by the hydrophilic substrate eventually freezes into a thin continuous ice sheet \cite{zhang2012condensation}. By employing a microscopy technique, the morphology of individual frost sites can be clearly visualized as shown in Fig.~\ref{fig:frost}(b). Most of the droplets are connected by ice bridges, and noticeable out-of-plane dendrite structure is not observed. We identified and calculated the size of each frost droplet and obtained a distribution as shown in Fig.~\ref{fig:frost}(c). The result suggests that the droplet size distribution peaks around the region of $A=1000~\rm{\mu } m^2$ and over 90\% of the frost droplet have an equivalent radius $r_{eqv}=\sqrt{A/\pi}$ less than 30 $\mu$m, which is far smaller than the length scale of the spreading droplet. Thus, the influence of discontinuity caused by this condensation frost on the spreading dynamics can be neglected.

Similar to the ice, the frost layer can remove the ice nucleation barrier \cite{lolla2022arrested} and provide a hydrophilic-like interface with a static contact angle measured as $\theta _e \approx 12 ^{\circ}$ \cite{knight1967contact}. Moreover, the thin frost layer can avoid the effect of substrate melting \cite{roisman2010fast} on the spreading dynamics. 

\subsection{Liquid properties}

The experimental Weber numbers are defined by $We=\rho D_0 U^2/\gamma$, where $\rho$ is the density of the liquid droplet, $D_0$ is the droplet diameter before impact, $U$ is the droplet impact velocity and $\gamma$ is surface tension. The liquid properties can vary with salinity, as shown in Table.~\ref{tab:my_label}. 

\begin{table}[htbp]
    \centering
    \tabcolsep=0.5cm
        \caption{Properties of the saline droplet}
         \label{tab:my_label}
    \begin{tabular}{ccccc}
   
    \hline
       Salinity  & Density & Freezing point & Surface tension & Thermal diffusivity\\
        $S_{i}$ & $\rho$ & $T_{f}$ & $\gamma$ & $\alpha$ \\
        wt$\%$ & kg/m$^{3}$ & $^{\circ}C$ & mN/m & m$^2/$s\\ \hline
        0 &  999.8&  0 & 72.2 & 1.42$\times 10^{-7}$\\\
        3.5 &  1028&  -2.1 & 73.4 & 1.39$\times 10^{-7}$\\
        10 &  1079&  -6.65 & 75.4 & 1.35$\times 10^{-7}$\\
        15 &  1118&  -11.05 & 76.9 & 1.33$\times 10^{-7}$\\
        20 &  1156&  -16.75 & 78.5 & 1.31$\times 10^{-7}$\\\hline
    \end{tabular}
\end{table}

\begin{figure*}[htbp]
\includegraphics[width=0.9\textwidth]{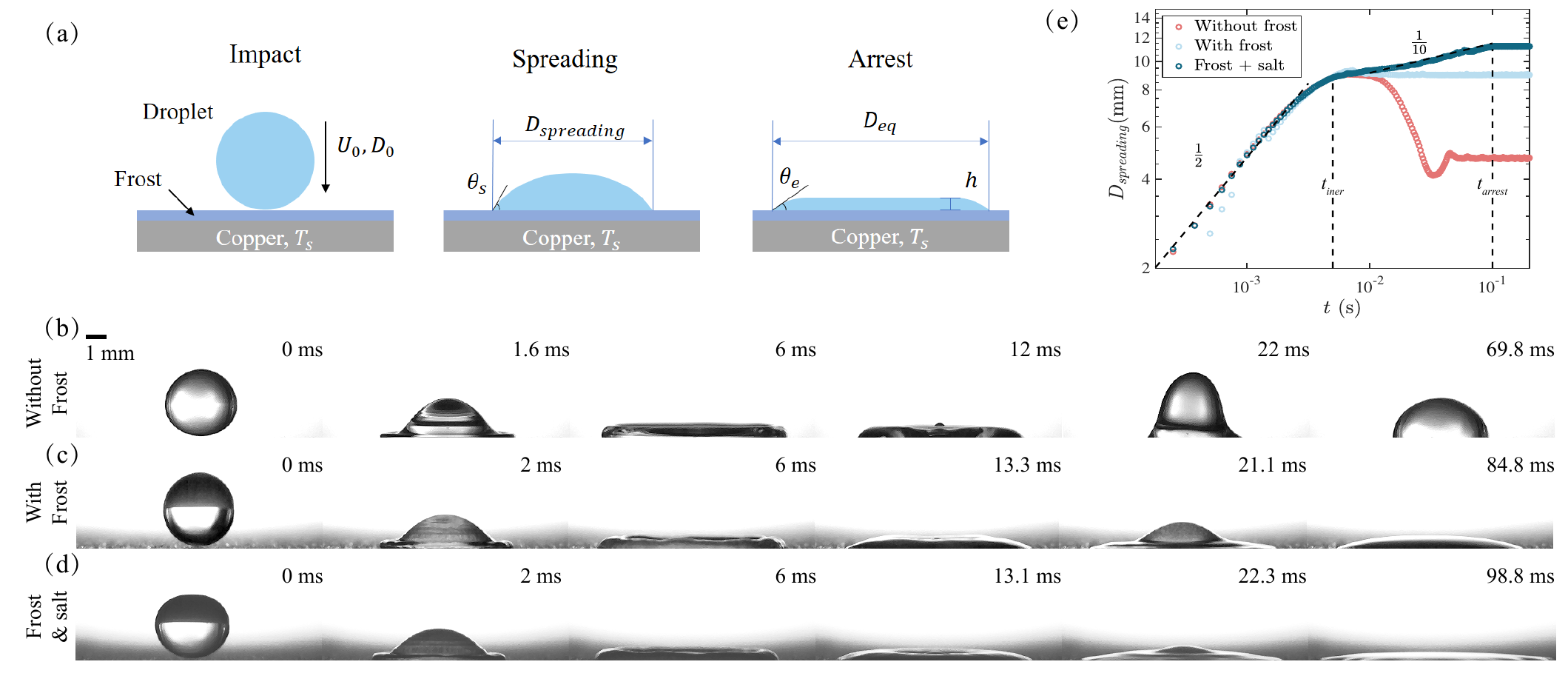} 
\caption{\label{fig:seq} Extended spreading dynamics of saline droplet impacting on a frosty substrate. (a) Schematic of droplet impacting, spreading and arresting. (b-d) Snapshots for water and saline droplets impacting on the cold substrate maintained at the same temperature $T_{s}=-20 ^{\circ}$C and the same velocity $U_{0}=1.0$ m/s: (b) Water droplet impacts on a surface without frost; (c) Water droplet impacts on a surface with frost; (d) Saline droplet ($S_{i}=15\%$) impacts on a surface with frost; (e) The spreading diameter versus time corresponds to the droplets in (b-d).}
\end{figure*}

\section{Result and discussion}
\subsection{Droplet spreading dynamics}
Fig.~\ref{fig:seq}(a) depicts the sketch of saline droplet impacting on a frosty substrate. The sequence of snapshots and the spreading diameter $D_{spreading}$ for droplets impacting on cold substrates are shown in Fig.~\ref{fig:seq}(b-e) under three typical conditions at the same substrate temperature $T_{s}=-20 ^{\circ}$C and impact velocity $U_{0}=1.0$ m/s. In all cases, the droplet initially spreads due to the impact inertia, which obeys the $1/2$ power law \cite{gordillo2018dynamics}. Without frost, as shown in Fig.~\ref{fig:seq}(b), due to the time delay between the droplet contact and ice nucleation \cite{kant2020fast}, the droplet retracts back to an equilibrium radius after reaching the maximal spreading, which is similar to the isothermal case \cite{laan2014maximum,josserand2016drop}. With the presence of a frost layer on the substrate, as shown in Fig.~\ref{fig:seq}(c), the moving contact line gets pinned at the maximal radius, which results from the growth of ice crystal \cite{grivet2022contact,koldeweij2021initial}. With an increased droplet salinity $S_{i}$, as shown in Fig.~\ref{fig:seq}(d), additional spreading occurs after the first inertia spreading phase upon impact. The arrest time for the moving contact line is significantly prolonged in this additional spreading phase, which further extends the spreading dynamics of the saline droplet. A transition from $1/2$ spreading power law to $1/10$ power law is observed. The $1/10$ exponent signifies a capillary-viscous spreading at the extended stage \cite{biance2004first}, which is hardly reported in previous studies for water droplets impact freezing \cite{lolla2022arrested,gordillo2018dynamics}.

\begin{figure*}[htbp]
\includegraphics[width=\textwidth]{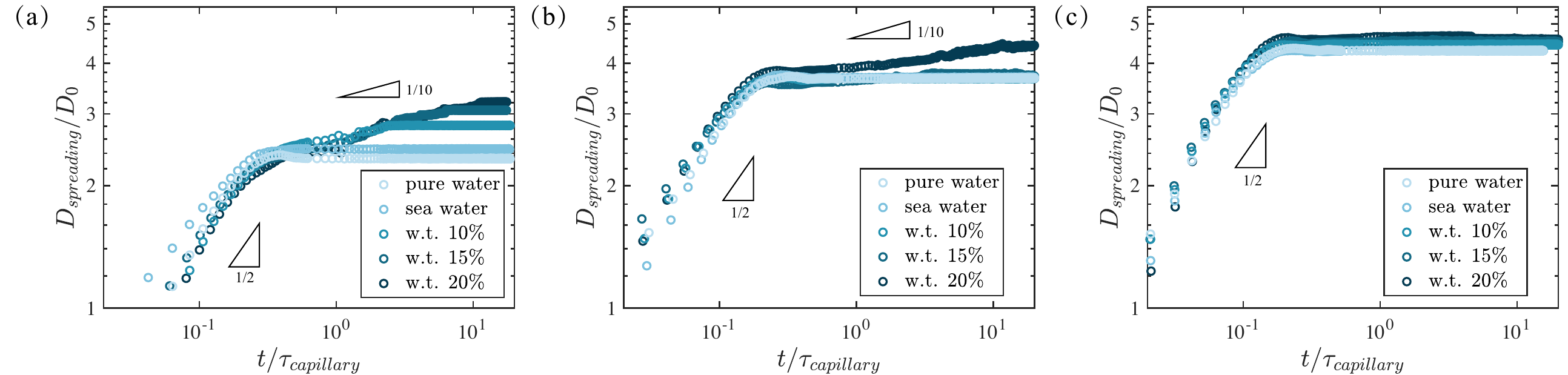}
\caption{\label{fig:spdcurves} Effect of $U_0$ and $S_i$ on the spreading dynamics of saline droplets impacting on a frosty substrate. (a) $U_{0}=0.8$ m/s, $We \approx 30$; (b) $U_{0}=1.9$ m/s, $We \approx 180$; (c) $U_{0}=2.9$ m/s, $We \approx 400$. The substrate is maintained at the same temperature $T_{s}=-15 ^{\circ}$C.}
\end{figure*}

Then, to investigate the effect of impact velocity $U_0$ and salinity $S_i$ on the spreading dynamics, we perform a series of experiments at various salinity [$S_{i} =$ 0, 3.5\% (typical salinity of sea water), 10\%, 15\%, 20\%] and various impact velocity, as shown in Fig.~\ref{fig:spdcurves}. For small impact velocity $U_0=0.8$ m/s, the final arrested diameter $D_{arrest}$ increases with the increase of salinity $S_i$ significantly, while the deviation tends to be negligible at large impact velocity $U_0=2.9$ m/s. The extended $1/10$ spreading regime is only observed at small impact velocity $U_0$ and high salinity $S_i$, while either large impact velocity $U_0$ or low salinity $S_i$ would lead to the arrest of contact line immediately after the $1/2$ inertial spreading stage.

\subsection{The maximum spreading diameter}
\begin{figure}[htbp]
\includegraphics[width=0.45\textwidth]{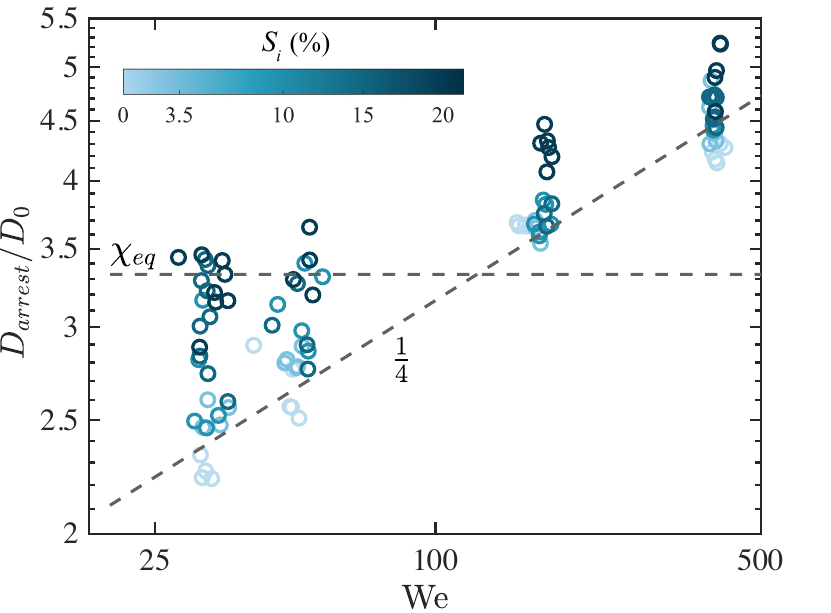} 
\caption{\label{fig:spdratio} Normalized arrested diameter $D_{arrest}/D_{0}$ as a function of $We$ of saline droplet impacting on a frosty substrate.}
\end{figure}

Figure~\ref{fig:spdratio} shows the dimensionless arrested diameter of droplet $D_{arrest}/D_0$ versus the Weber number $We$ with various salinity $S_i$. For pure water droplets, multiple scaling relations have been proposed to predict the maximal spreading of droplet impacting, either isothermal case \cite{clanet2004maximal,gordillo2018dynamics} or freezing case \cite{lolla2022arrested,kant2020fast,thievenaz2020freezing,hu2020frozen}, depending on the experimental conditions. In our experiments, for dilute saline droplets at large impact velocity $U_0$, the final arrested diameter $D_{arrest}$ is found to be well described by the $1/4$ scaling law $D_{arrest} \sim D_0 We^{1/4}$ 
\cite{clanet2004maximal}, where $We=\rho D_0 U_0^2/\gamma$, $\rho$ and $\gamma$ are the density and surface tension of the droplet. 

\begin{figure*}[htbp]
\includegraphics[width=0.5\textwidth]{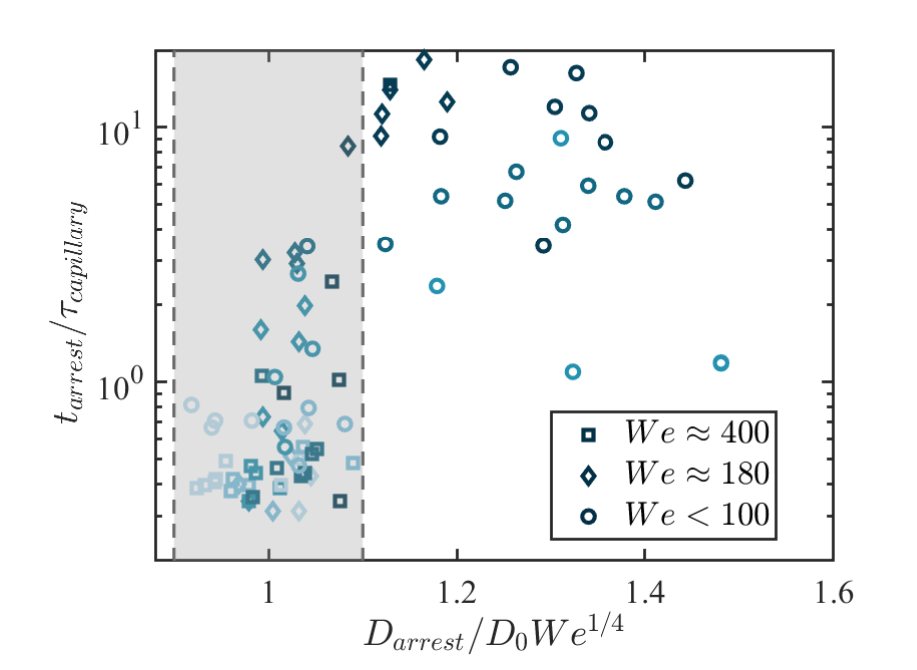} 
\caption{\label{fig:time}  Normalized arrest time of saline droplet impacting on a frosty surface.}
\end{figure*}

However, for small Weber number $We<100$ and high salinity $S_i$ cases, the arrest time $t_{arrest}$ is significantly prolonged by the extended spreading stage, leading to an increase in droplet arrest diameter $D_{arrest}$ of up to 40\%, as shown in Fig.~\ref{fig:time}. The droplet's arrest time $t_{arrest}$ is normalized by the capillary timescale $\tau_{capillary}$, and the dimensionless arrested diameter $D_{arrest}/D_{0}$ is further normalized by the $We^{1/4}$ scaling law to estimate the contribution from the additional capillary spreading. For dilute saline droplet at large Weber number ($We>100$), the droplet's arrest time is comparable to the capillary time scale, and the final arrest diameter agrees well with the isothermal scaling law, with a deviation less than 10\% as indicated by the shadow region. However, for droplets with low impact velocity ($We < 100$) and high salinity, the arrest time  significantly exceeds the capillary timescale, allowing the occurrence of additional spreading.

To further quantify the contribution of the secondary spreading stage, we first focus on the criterion for the onset of the extended capillary-viscous regime. From the perspective of wetting, as illustrated in Fig.~\ref{fig:seq}(a), when the dynamic contact angle at the spreading front, $\theta_s$, is greater than the equilibrium contact angle on the frosty substrate, $\theta_e$, the capillary force drives the front to advance. It eventually comes to a halt upon reaching an equilibrium state where $\theta_s=\theta_e$ \cite{gennes2004capillarity}.
By considering volume conservation of the droplet, the diameter for the droplet at equilibrium state $D_{eq}$ can be obtained for a given initial volume $V_0$ (details in Appendix). This equilibrium spreading ratio, $\chi_{eq}=D_{eq}/D_0=3.328$, serves as an upper limit for the arrested diameter of droplets $D_{arrest}$ with small impact velocity $U_0$ and high salinity $S_i$, as indicated by the horizontal dashed line in Fig.~\ref{fig:spdratio}, which agrees well with the experimental data.


\subsection{The spreading phase diagram}
From a cooling perspective, as the droplet's bulk temperature decreases from room temperature to a critical temperature, determined by the influence of kinetic undercooling \cite{de2017contact,lolla2022arrested}, the contact line arrests.
Thus, the onset of the extended spreading stage requires a longer cooling timescale $\tau _{cooling}$ than the spreading timescale $\tau _{capillary} \sim \sqrt{\rho D_0^3/ \gamma}$. By considering the balance between the heat conduction from the substrate and sensible heat of the droplet, we obtain the cooling timescale $\tau _{cooling}$ \cite{lolla2022arrested}:
\begin{equation}
    \label{eq:t_cooling}
    \tau_{cooling} \sim \frac{T_{0}-T_{f}}{\Delta T} \frac{h^{2}}{\alpha},
\end{equation}
where $\alpha$ is the thermal diffusivity of the droplet, $\Delta T=T_f-T_s$ is the substrate supercooling temperature, which decreases with the increase of salinity $S_i$ due to the freezing point depression. $T_0$ and $T_f$ is the initial temperature and freezing temperature of the droplet, respectively. Here, $h$ is average thickness of the spreading liquid film, which can be estimated by considering volume conservation $h\sim D_0^3/D^2$ at the moment when the droplet reaches the maximal inertial spreading $D \sim D_0We^{1/4}$. 

The onset criterion of the extended regime can be deduced by equating $\tau _{cooling}=\tau _{capillary}$, and the critical Weber number $We_{cr}$ for the spreading transition at a given supercooling temperature $\Delta T$ can be expressed as:
\begin{equation}
\label{eq:relation}
We_{cr} \sim C(\Delta T_i/\Delta T-1),
\end{equation}
where $C=(D_0\gamma /\rho)^{1/2}/\alpha$, $\Delta T=T_f-T_s$ is the supercooling temperature which typically decreases with the increase of salinity $S_i$, $\Delta T_i=T_0-T_s$ is the difference between the droplet initial temperature and the substrate temperature, which is independent on the freezing point of the droplet. 
\begin{figure}[htbp]
\centering
\includegraphics[width=0.45\textwidth]{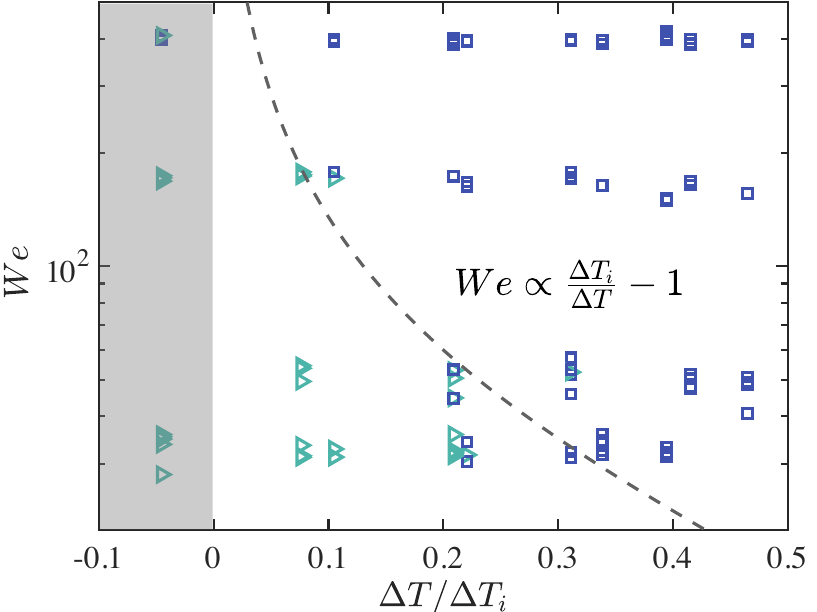} 
\caption{\label{fig:phase_diagram} Phase diagram for spreading dynamics of saline droplet impacting on a frosty substrate. The triangle and the square represent the presence and absence of the extended spreading stage, respectively. The shaded region represents the spreading process without freezing.}
\end{figure}

Equation~\ref{eq:relation} reveals the competition between dynamic spreading $We$ and bulk cooling efficiency $\Delta T/\Delta T_i$ for droplets with different salinity $S_i$, which agrees well with the experimental phase diagram for spreading dynamics of saline droplet impacting on a frosty substrate, as indicated by the dash line in Fig.~\ref{fig:phase_diagram}. For the region below the dash line ($We<We_{cr}$), corresponding to $\tau _{capillary}<\tau_{cooling}$, the contribution of the extended spreading stage must be taken into account. The critical supercooling temperature $\Delta T$, corresponding to the critical salinity $S_i$, decreases rapidly with the increase Weber number and it approaches zero for high Weber numbers. This explains why the $1/10$ spreading power law has hardly been reported in previous studies on water droplet impacting and freezing.

\subsection{Scaling analysis on the salinity effect}
To further quantify the arrested condition during the extended spreading stage, a scaling analysis is conducted by integrating the well-known Tanner's law for capillary-viscous spreading with the kinetic freezing theory \cite{tanner1979spreading,de2017contact}. In this stage, the evolution of spreading diameter gives:
\begin{equation}
    \label{eq:tanner}
    D_{spreading} \sim D_0(t/\tau_{viscous})^{1/10},
\end{equation}
where $\tau_{viscous}=\mu D_{0}/\gamma$ is the viscous timescale, $\mu$ is the dynamic viscosity. Then, the contact line velocity can be obtained by $v_{cl}\sim \partial{D_{spreading}}/\partial{t}$. 
For the arresting of a moving contact line, the freezing front velocity $v_{front}$ must catch up with the velocity of contact line to advance \cite{de2017contact}. The freezing front velocity can be expressed as: $v_{front}=\kappa \Delta T^b$, where $b=2$ is a non-dimensional parameter and $\kappa=0.0028$ m/(s $\cdot$ K$^{2}$) is the kinetic cooling coefficient for water droplet on a icy substrate \cite{lolla2022arrested}.

Equating the freezing front velocity $v_{front}$ and the contact line velocity $v_{cl}$, the arrest timescale is then derived:
 \begin{equation}
     \label{eq:t_arrest}
     t_{arrest} \sim D_0 (\gamma/\mu)^{1/9} (\kappa \Delta T^b )^{-10/9}.
 \end{equation}

Inserting Eq.~\ref{eq:t_arrest} into Eq.~\ref{eq:tanner}, at the extended capillary-viscous regime, we find the arrest diameter $D_{arrest}$ can be expressed by a $-2/9$ power law as:
\begin{equation}
    \label{eq:w_cap}
    D_{arrest}/D_0 \sim  (\Delta T/\sqrt{\gamma/\mu \kappa})^{-2/9},
\end{equation}
where $\sqrt{\gamma/\mu \kappa}$ combines the physical parameter related to the liquid property and kinetic freezing theory.

\begin{figure}[htbp]
\centering
\includegraphics[width=0.46\textwidth]{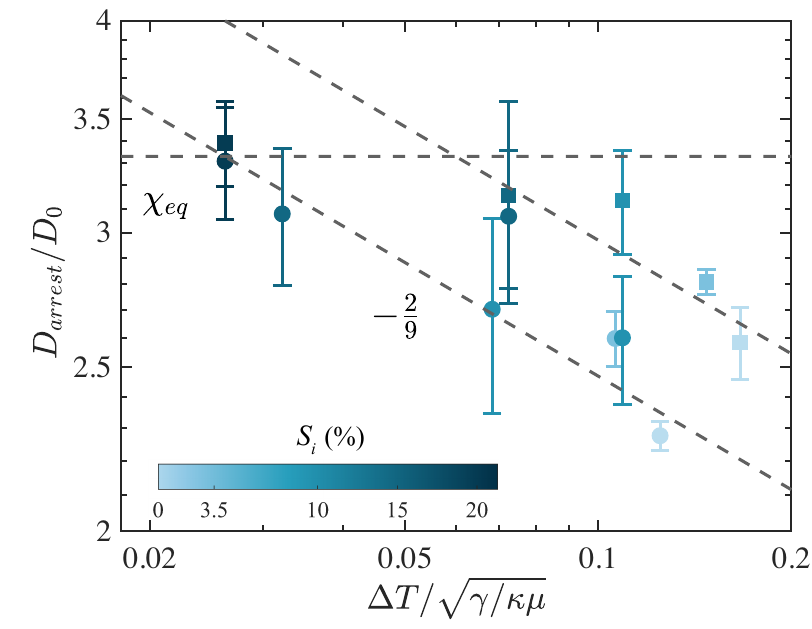} 
\caption{\label{fig:scaling} Dimensionless arrested diameter $D_{arrest}/D_{0}$ as a function of dimensionless supercooling temperature $\Delta T/ \sqrt{\gamma / \kappa \mu}$. The square and the circle represent $We \approx 80$ and $We \approx 30$, respectively.}
\end{figure}

As shown in Fig.~\ref{fig:scaling}, the derived $-2/9$ power law collapses all of the experimental data for $We<100$ and $\Delta T>0$ cases, and the upper limit at high salinity $S_i$ also agrees well with the derived equilibrium spreading ratio $\chi _{eq}=D_{eq}/D_0$.

Based on the above understandings, we propose a unified prediction model for the final arrested diameter $D_{arrest}$ of saline droplet impacting on a frosty substrate. As shown in Fig.~\ref{fig:setup}(e), the final arrested diameter can be divided into two parts contributed by inertia and capillarity, respectively:
\begin{equation}
\label{eq:D_arrest}
    D_{arrest}=D_{inertia}+\delta D_{capillary},
\end{equation}
where $D_{inertia}$ is the maximal spreading diameter attained by the inertial spreading which can be estimated by the $1/4$ power law \cite{clanet2004maximal}, $D_{capillary}$ corresponds to the distance that the contact line advances in the extended capillary-viscous spreading regime, and $\delta$ is a non-dimensional parameter that describes the status of droplet prior to arrest. 

\begin{figure}[htbp]
\centering
\includegraphics[width=0.44\textwidth]{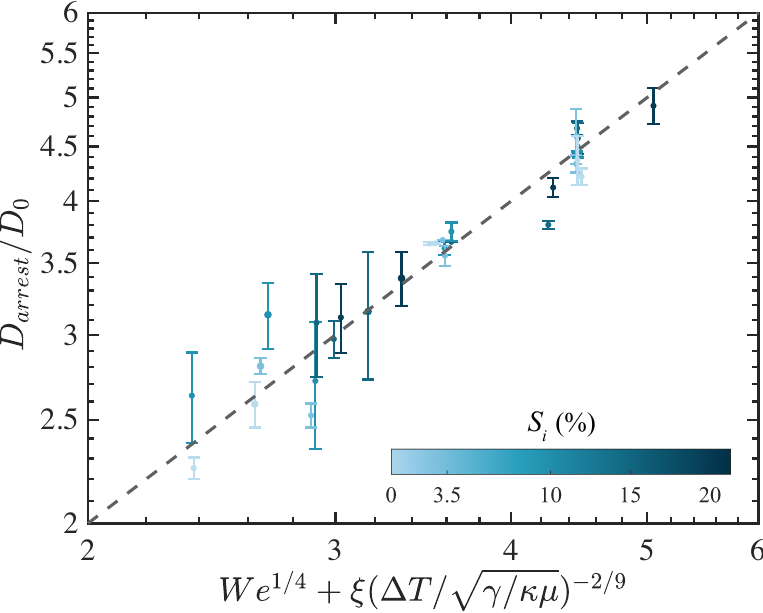} 
\caption{\label{fig:prediction} Comparison of the dimensionless arrested diameter $D_{arrest}/D_{0}$ between the experimental measurements and the model predictions.}
\end{figure}

When $We < We_{cr}$, corresponding to the region below the dash line in Fig.~\ref{fig:phase_diagram}, with the presence of extended spreading regime, $\delta=1$. On the other hand, when $We > We_{cr}$, with the absence of extended spreading regime, $\delta=0$. Thus, by equating Eq.~\ref{eq:w_cap} and Eq.~\ref{eq:D_arrest}, the proposed prediction model on the final arrested diameter can be expressed as:
\begin{equation}
\label{eq:model}
    D_{arrest}/D_0=We^{1/4}+\xi (\Delta T/\sqrt{\gamma/\mu \kappa})^{-2/9},
\end{equation} 
where $\xi = \delta \epsilon $, $\epsilon$ is a numerical parameter determined from fitting of the experimental data to give a best fit of $\xi=0.29$ when $We < We_{cr}$. The supercooling temperature $\Delta T$ considers the depression of the freezing point induced by salinity $S_i$. Inserting the prior known physical parameters into Eq.~\ref{eq:model}, our model can provide a good estimation of the final arrested diameter $D_{arrest}$ for saline droplets impacting on a frosty substrate over a wide range of $We$ and $S_i$ as shown in Fig.~\ref{fig:prediction}.

In summary, the presence of frost and salt may significantly affect the spreading and arresting dynamics for droplets impact upon a cold substrate. By performing the controllable and reproducible experiments of saline droplets impacting on a frosty substrate, we highlight the extended $1/10$ capillary-viscous spreading regime except for the well-studied $1/2$ inertial spreading regime. The onset criterion and the arrested condition are derived from both the wetting and cooling perspectives. By incorporating the effect of kinetic undercooling into the classical Tanner's law, we establish a $-2/9$ scaling law to characterize the droplet arrest diameter $D_{arrest}$ against supercooling degree $\Delta T$, and propse a unified model to predict the final arrested diameter across a wide range of Weber numbers ($We$) and initial salinities ($S_i$). Our findings provide a comprehensive insight for a saline droplet impacting and freezing process, involving the impact inertia, the salinity, the substrate temperature and the frost. Our results offer new insights and guidance for understanding and predicting icing phenomena in various applications, such as ice accretion on marine vessels and offshore structures \cite{dehghani2017sea,ryerson2011ice,lam2020effect}.

\begin{acknowledgments}
This work has been supported by the Natural Science Foundation of China under Grant No. 11988102, the National Key R\&D Program of China under Grant No. 2021YFA0716201, the New Cornerstone Science Foundation through the New Cornerstone Investigator Program and the XPLORER PRIZE, and Shuimu Tsinghua Scholar Program No. 2023SM038. We are grateful to Mingbo Li, Sijia Lyu, and Lei Yi for helpful discussions. 
\end{acknowledgments}

\appendix
\section{Physical Model for Equilibrium diameter}
When a droplet is placed on a partially wetting plane, the free liquid surface is actually determined by balancing of surface tension and gravity. In particular, the droplet shape would change from a sphere cap to a flat pancake with an increasing volume. 
 \begin{figure*}[htbp]
\includegraphics[width=0.5\textwidth]{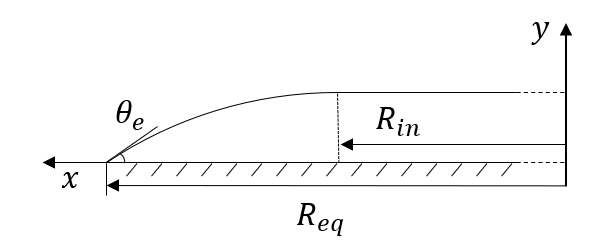} 
\caption{\label{fig:radius} Sketch of the droplet profile.}
\end{figure*}

For small droplets with the equilibrium radius $R_{eq}$ smaller than the capillary length $\kappa^{-1}$, the droplet shape is only determined by the surface tension. The droplet is shaped like a sphere cap whose edges intersect the substrate at angle $\theta_{e}$. Applying volume conservation $V_0=V_{eq}$ to the liquid, we obtain:
\begin{equation}
    \frac{\pi}{6}D_0^3=\frac{\pi}{3}R_{eq}^3\frac{2-3\text{cos}\theta_{e}+2\text{cos}^2\theta_{e}}{\text{sin}^3\theta_{e}}.
\end{equation}

For large droplets whose equilibrium radius is greater than the capillary length,  gravity dominates over the surface tension. As shown in Fig.~\ref{fig:radius}, the droplet is flattened to take the shape of a liquid pancake. The droplet profile at edge can be easily calculated by equating the surface force and hydro-static pressure in the horizontal direction \cite{gennes2004capillarity}. In our case, the equilibrium contact angle $\theta_e=12^\circ$ \cite{knight1967contact,thievenaz2020retraction} is small, thus the curve can be approximated as a portion of sphere cap with a radius of curvature $r$. And the difference between the radius of inner cylinder $R_{in}$ and the equilibrium radius can be approximated as the capillary length, this gives:
\begin{equation}
    \label{eq:r_in}
    \kappa^{-1}=R_{eq}-R_{in}.
\end{equation}

Integrating the droplet profile along the axis of rotational symmetry and applying the volume conservation,  we are able to obtain a relation:
\begin{equation}
    \label{eq:volume2}
    \frac{\pi}{6}D_0^3=\frac{r}{12}\left [r^2(-9\text{cos}\theta_e+\text{cos}3\theta_e+8)-6rR_{in}(\text{sin}2\theta_e-2\theta_e)-12R^2_{in}(\text{cos}\theta_e-1)\right ].
\end{equation}

Combining Eq.~\ref{eq:r_in} and Eq.~\ref{eq:volume2}, and plugging in the droplet initial diameter $D_0=3.4$ mm, we are able to obtain the equilibrium spreading ratio as: $\chi_{eq}=D_{eq}/D_0=3.328$.


%

\end{document}